\documentclass[
  aps,
  prx,
  preprint,
  tightenlines,
  superscriptaddress,
  amsmath,
  amssymb,
  floatfix,
]{revtex4-2}

\usepackage[T1]{fontenc}
\usepackage{lmodern}
\usepackage{graphicx}
\usepackage{booktabs}
\usepackage{multirow}
\usepackage{longtable}
\usepackage{xurl}
\usepackage{microtype}
\usepackage{xcolor}
\usepackage{hyperref}
\usepackage{etoolbox}
\usepackage{xr-hyper}

\hypersetup{
  colorlinks=true,
  linkcolor=black,
  citecolor=black,
  urlcolor=blue!55!black,
}

\makeatletter
\patchcmd{\frontmatter@abstract@produce}{%
  \preprintsty@sw{%
   \do@output@MVL{%
    \vskip\frontmatter@preabstractspace
    \vskip200\p@\@plus1fil
    \penalty-200\relax
    \vskip-200\p@\@plus-1fil
   }%
  }{%
   \addvspace{\frontmatter@preabstractspace}%
  }%
}{%
  \addvspace{0.7\baselineskip}%
}{\typeout{abstract-page-break patch applied}}{\typeout{abstract-page-break patch FAILED}}
\def\frontmatter@postabstractspace{0.3\baselineskip}
\def\selectlanguage{\@gobble}
\newcommand*{\addFileDependency}[1]{
  \typeout{(#1)}
  \@addtofilelist{#1}
  \IfFileExists{#1}{}{\typeout{No file #1.}}
}
\makeatother

\newcommand*{\myexternaldocument}[1]{%
    \externaldocument[][nocite]{#1}%
    \addFileDependency{#1.tex}%
    \addFileDependency{#1.aux}%
}
\myexternaldocument{supplementary_material}

\begin{document}

\title{Structure alone supports efficient visual computation in the \emph{Drosophila} visual system}

\author{Eudald Correig-Fraga}
\email[Contact author: ]{eudald.correig@innovamat.com}
\affiliation{Innovamat Education, Sant Cugat del Vallès, Catalonia}
\affiliation{Department of Chemical Engineering, Universitat Rovira i Virgili, Tarragona, Catalonia}

\author{Roger Guimerà}
\email[Contact author: ]{roger.guimera@urv.cat}
\affiliation{Department of Chemical Engineering, Universitat Rovira i Virgili, Tarragona, Catalonia}
\affiliation{Center for Computational Science and Applied Mathematics (ComSCIAM), Universitat Rovira i Virgili, Tarragona, Catalonia}
\affiliation{ICREA, Barcelona, Catalonia}

\author{Marta Sales-Pardo}
\email[Contact author: ]{marta.sales@urv.cat}
\affiliation{Department of Chemical Engineering, Universitat Rovira i Virgili, Tarragona, Catalonia}
\affiliation{Center for Computational Science and Applied Mathematics (ComSCIAM), Universitat Rovira i Virgili, Tarragona, Catalonia}

\date{\today}

\begin{abstract}
Understanding the extent to which measured synaptic wiring determines computation remains a central challenge. Here, we couple the proofread adult \emph{Drosophila melanogaster} connectome to an anatomically faithful model of its eye. Visual information is inputted in the eye model, then passed to the connectome, and finally read from a Kenyon-cell-centered linear decoder. This creates a connectome‑only model in which the anatomical graph and eye geometry are fixed and only scalar synaptic gains and neuronal thresholds may be learned. The model supports multitask vision, including color discrimination, shape classification, and numerical discrimination that follows a ratio-dependent scaling characteristic of approximate number perception. To test whether precise connectivity is consequential under wiring economy, we compare the biological graph to randomized ensembles that increasingly preserve biological synaptic constraints. At matched wiring cost, the biological network consistently yields higher accuracy, whereas less constrained rewiring surpasses it at the cost of inflated wiring. These findings indicate that the measured connectivity and eye geometry jointly set efficient operating points for visual computation.
\end{abstract}

\maketitle
\enlargethispage{-16pt}

\section{Introduction}

Deciphering how physical connectivity shapes neural computation is a foundational goal of neuroscience. The recent availability of full adult connectomes for {\it Drosophila melanogaster} represents a qualitative leap in our ability to address this challenge, offering the first complete wiring diagram of a brain capable of complex behaviors~\cite{scheffer_connectome_2021, atanas_brain-wide_2023, dorkenwald_neuronal_2024, schlegel_whole-brain_2024, creamer_bridging_2024}. While the connectome far from captures the full range of biological dynamics, it is a powerful generator of testable hypotheses~\cite{jabr_connectome_2012} and can help us answer very fundamental questions regarding the computational capabilities of biological brains. One of these questions is to find the principles of design behind brain architectures that allow individuals to perform multiple complex tasks with a limited set of neurons. From an evolutionary perspective, we anticipate that these circuits have been optimized for performance under strict biophysical constraints, particularly those associated with energetic costs~\cite{chklovskii_wiring_2002, laughlin_energy_2001, bullmore_economy_2012, mitchison_neuronal_1991}.

To investigate how wiring economy shapes computation, we built a computational model of the fruit fly brain, in which information flow is dictated solely by the measured synaptic wiring. We construct a message-passing network based on the graph provided by the proofread adult \emph{Drosophila} connectome  (Fig.~\ref{fig:figure_1}a). By coupling this system to a realistic model of the compound eye~\cite{nern_connectome-driven_2025}, we train the network on visual tasks ranging from color and shape recognition to numerosity discrimination. We then compare its performance to ensembles of randomized networks that preserve the number of neurons and synapses while differing in wiring constraints that span a spectrum of biological plausibility, from unconstrained random graphs to anatomically constrained variants that match biological restrictions of a neural connectome. This design allows us to test to what extent the measured wiring alone can support non-trivial visual computations. Our results indicate that the biological connectome occupies a privileged position within the space of wiring-economical architectures defined by these randomizations, as it can optimally perform visual tasks within that ensemble. 

\section{Results}

\subsection{A connectome-only model to assess visual task performance}

\begin{figure*}[p]
\centering
\includegraphics[width=0.62\textwidth]{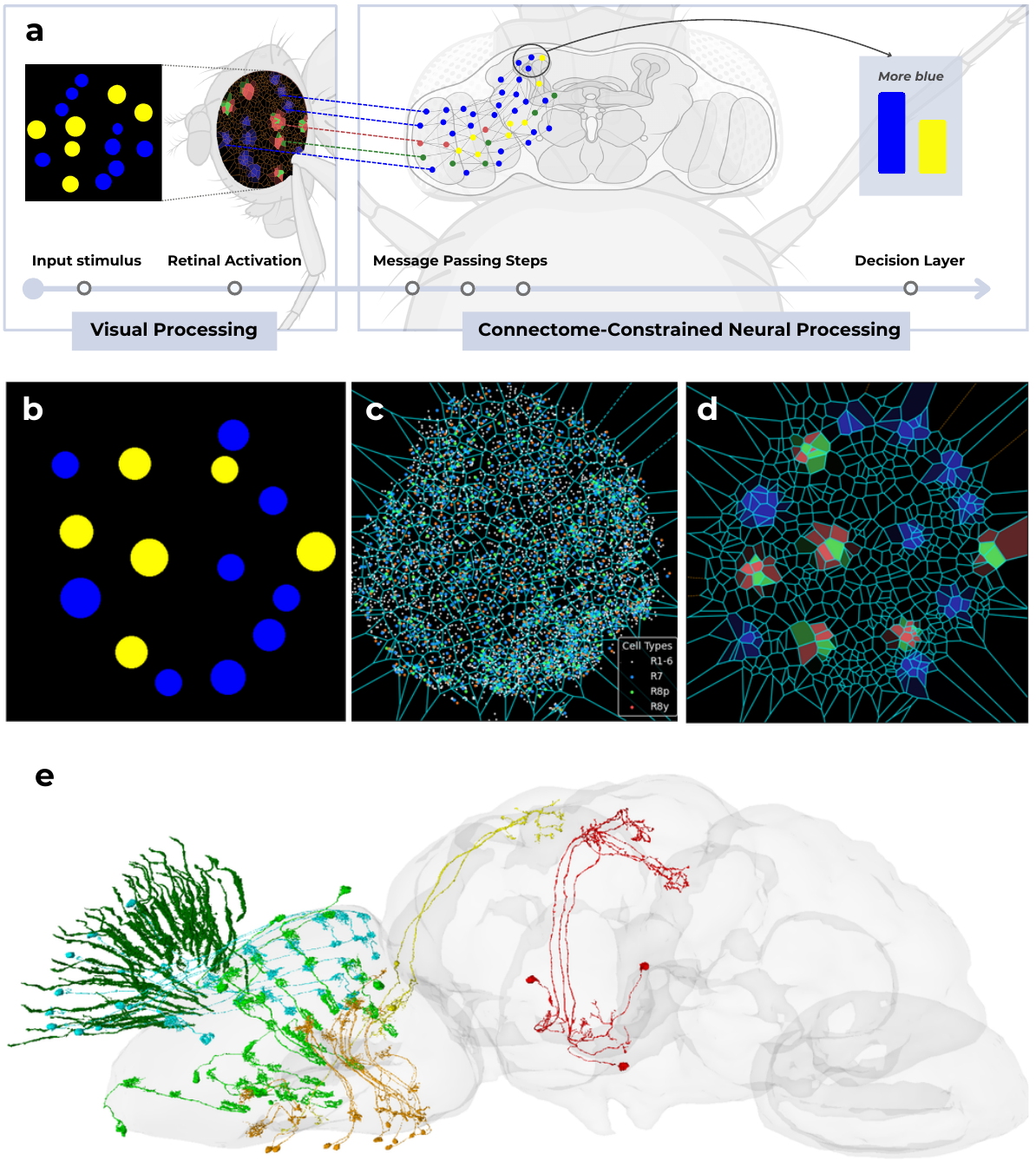}
\caption{\small\textbf{Computational modeling of numerical discrimination in Drosophila melanogaster.} {\bf a,} Visual stimuli with varying numerosities of colored dots (blue vs. yellow) are processed through an anatomically accurate model of the fly's compound eye. We reconstruct the 3D positions of retinal neurons from the connectome data, project them onto a 2D plane, and use Voronoi tessellation to map ommatidia, with R7 neuron positions as cell centers. Each ommatidium samples its corresponding region of the visual field, with specific photoreceptors (R1-R8) encoding different color channels based on their biological spectral sensitivities. The captured light intensity in each retinal neuron produces activation patterns that form the initial input to the neural network. These retinal neuron activations propagate via a message-passing algorithm on the graph created by the \emph{Drosophila} connectome (54.5 million machine-generated synaptic contacts). Information flows via message passing for 2-6 propagation steps~\cite{SupplementalMaterial}, mimicking how neural signals travel from the retina through the optic lobe and ultimately to the Kenyon cells in the mushroom body, which serve as the decision-making locus. \textbf{b,} Example of an input image. \textbf{c,} Distribution of photoreceptor terminals colored by their spectral sensitivity (R1-6: white, broad spectrum; R7: blue, UV-sensitive; R8p: green, green-sensitive; R8y: red, red-sensitive). The Voronoi tessellation (turquoise) shows the spatial organization of these terminals. \textbf{d,} Pattern of neuronal activation in response to an example stimulus in \textbf{b}. The yellow dots activate green and red receptors (R8p and R8y, respectively) in the appropriate Voronoi cells; the blue dots activate only the blue (R7) receptors. \textbf{e} Biologically accurate representation of a sample of 100 neurons in the areas of interest of the connectome (left hemisphere), with retina cells in dark green, lamina in blue, medulla in lime, lobula in yellow, and lobula plate in orange. We also show mushroom body KCapbp-m (decision-making) cells in red for reference. Image \textbf{e} was created with~\url{https://codex.flywire.ai/}}
\label{fig:figure_1}
\end{figure*}
We build a model of the eye by inferring its ommatidial structure from the positions of the retinal neurons from the connectome~\cite{dorkenwald_neuronal_2024, schlegel_whole-brain_2024}. This structure provides a principled way of tessellating the input image into tiles to match each ommatidium, thus ensuring that visual inputs have the correct angular and spectral resolution~\cite{wernet_building_2004, paulk_vision_2013, rister_dissection_2007, schnaitmann_color_2018} (Fig.~\ref{fig:figure_1}a-d; details in Methods). In this way, the luminosity in input images can be translated into activations of the retinal neurons of the connectome. Then, the input signal flows through the brain via a message-passing algorithm where, at each discrete step $k$, neurons aggregate presynaptic signals by a sum operator weighted by synapse-derived efficacies, apply a nonlinear activation function, and pass the updated state forward (Fig.~\ref{fig:figure_1}e and Eqs.~\eqref{eq:methods_message_passing1}-~\eqref{eq:methods_message_passing3} in Methods). Activity is then read out from the population of Kenyon cells in the mushroom body (Fig.~\ref{fig:figure_1}a,e). Although action selection in \textit{Drosophila} depends on distributed interactions beyond the Kenyon cells themselves~\cite{aso_mushroom_2014, shiu_drosophila_2024}, we use Kenyon-cell activity here as a simple, standardized readout layer for the model.

This setup makes our model “connectome-only”, in the sense that the anatomical graph and eye geometry are fixed, and no additional computational units are introduced. During learning, the only trainable connectome parameters are scalar gains that multiply the observed synapse counts on existing edges and are bounded in magnitude by one, so the resulting weights cannot exceed those counts in magnitude. They mimic biological mechanisms in which synapse strength is adapted through Hebbian learning~\cite{hebb_organization_2012,SupplementalMaterial}. This design contrasts with connectome‑constrained models that add hidden processing stages or task‑specific modules~\cite{lappalainen_connectome_constrained_2023, chen_data_based_2022, shiu_drosophila_2024} and lets us test the computational consequences of observed wiring and eye geometry alone.

\subsection{Randomized connectome ensembles to explore the relationship between wiring and task performance }

To explore the relationship between physical connectivity and brain function, we investigate whether randomizing the connectome to various degrees would change its performance in different visual tasks.  To that end, we define several randomization strategies that differ in their degree of biological plausibility according to wiring-economy principles~\cite{chklovskii_wiring_2002, bullmore_economy_2012}. Specifically, we use the sum over connected pairs of neurons of their Euclidean somatic distance multiplied by synapse count as a proxy for wiring cost (Methods), and then define four random connectome ensembles generated from the biological connectome and differing in the constraints we impose on the total wiring cost
(Fig.~\ref{fig:figure_2}a-f, and Supplemental Material~\cite{SupplementalMaterial}; Methods for details):
\begin{figure*}[p]
\centering
\includegraphics[width=\textwidth]{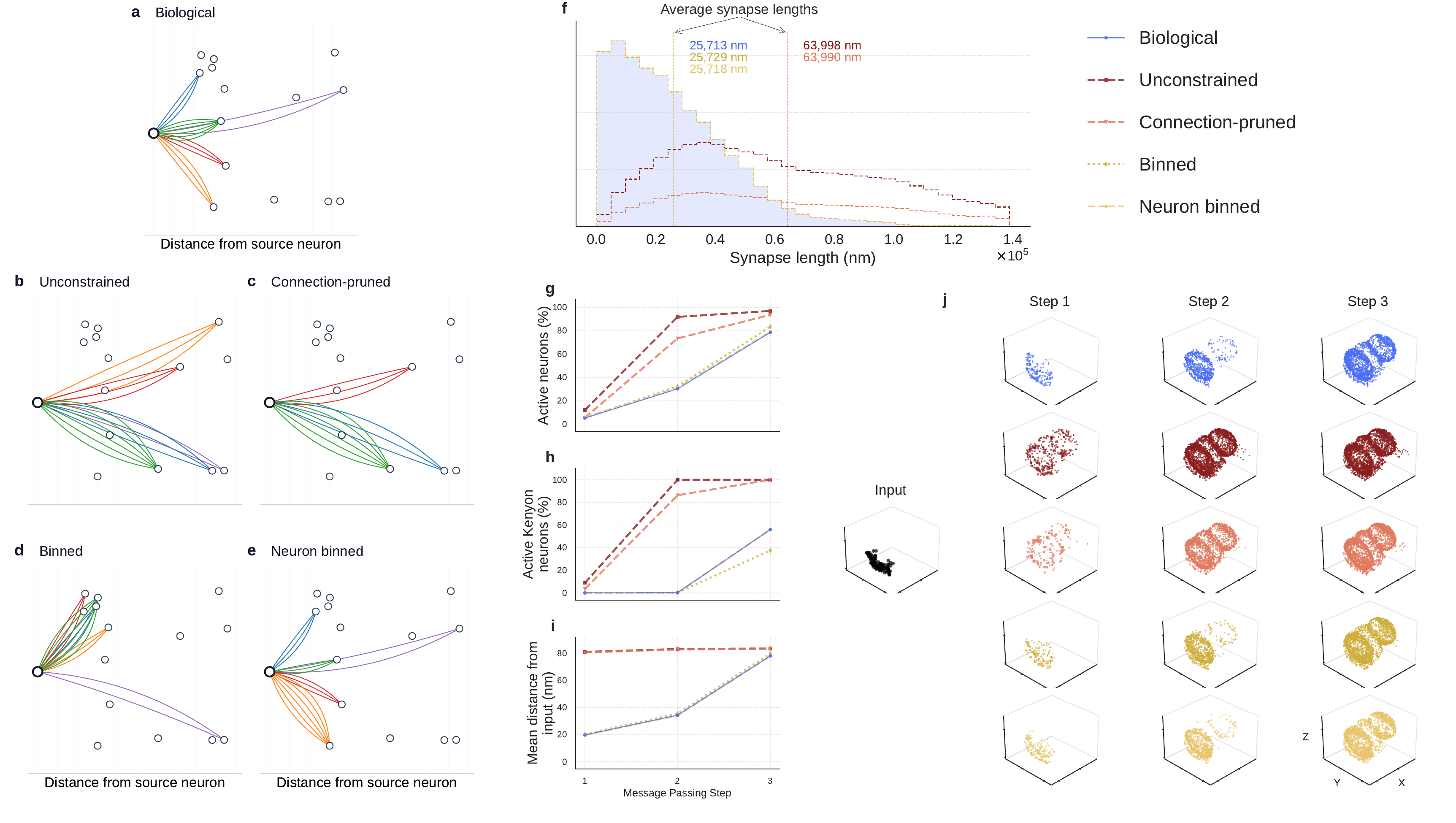}
\caption{\small
\textbf{Connectome randomization strategies and statistics.}
\textbf{a--e,} Toy illustrations of the five connectome ensembles.
\textbf{a,} Biological connectome.
\textbf{b,} Unconstrained randomization, where each neuron keeps the same number of outgoing connections, but targets are randomly shuffled. This produces networks with much larger total wiring and longer synapses on average than the biological connectome.
\textbf{c,} Connection-pruned randomization, obtained by starting from the unconstrained network and then removing connections until the total wiring matches the biological one. Even after this pruning step, the mean connection length remains above biological values.
\textbf{d,} Synapse-bin randomization, where neuron pairs are grouped by Euclidean soma-to-soma distance into 100 bins and synapses are shuffled only within the same bin. This keeps the global distribution of connection lengths, and therefore also preserves both the average synapse length and the total wiring budget.
\textbf{e,} Neuron-bin randomization, where for each neuron its outgoing synapses are divided into 20 bins according to wiring length, and randomization is restricted to each bin separately. In this ensemble, the network wiring pattern is unchanged, and only the synaptic weights between connected neuron pairs are redistributed.
\textbf{f,} Distribution of the synapse lengths for the five ensembles. The least constrained randomizations have much longer average lengths than the biological and the biologically constrained ones.
\textbf{g, h,} Fraction of neurons (\textbf{g}) and Kenyon neurons (\textbf{h}) active after each message-passing step ($N=3$). In these two plots we see that the least constrained networks very rapidly saturate the connectome, whereas the constrained ones, even though they reach large amounts $(\sim80\%)$ of neurons in very few steps, are more constrained in the amount of neurons to use for computation.
\textbf{i,} Mean distance of activated neurons after each step, which shows how the information travels very far very rapidly for unconstrained networks.
\textbf{j, } Propagation patterns for the biological and random ensembles. In the plots, we show one out of every hundred active neurons in their 3D position given by the connectome. We can see how, in biological and biologically constrained ensembles, the information slowly moves from the eye to the rest of the brain, whereas in the unconstrained ones the whole brain is almost immediately active.
}
\label{fig:figure_2}
\end{figure*}

\paragraph {Unconstrained ensemble.} For each neuron, we rewire outgoing connections at random while preserving per‑neuron out‑degree and ignoring synaptic length, leading to both total wiring and average synapse lengths of the resulting networks much larger than those of the biological connectome (Fig.~\ref{fig:figure_2}b,f).
    
\paragraph{Connection‑pruned ensemble.} We start from a network in the unconstrained ensemble, and then remove synaptic connections at random until the total wiring length equals that of the biological connectome. In this ensemble, the average edge length is higher than that of the biological connectome (Fig.~\ref{fig:figure_2}c,f).

\paragraph{Synapse‑bin ensemble:} Starting from the biological connectome, we group neuron pairs by Euclidean somatic distance in 100 distance bins. We then shuffle synapses only within the same bin, preserving the global edge‑length distribution (and thus average length) and the total wiring budget (Fig.~\ref{fig:figure_2}d,f).
    
\paragraph{Neuron‑bin ensemble.} For each neuron, we group its synapses in 20 bins depending on their wiring length, and then we randomize synapses only within each bin of each neuron. This results in networks in which neurons preserve their neighbors but with a different number of synapses between them (Fig.~\ref{fig:figure_2}e,f). 

\bigskip 
We find that differences in wiring constraints also result in differences in information propagation (Fig.~\ref{fig:figure_2}g–j). In the least constrained ensembles, activity rapidly saturates the connectome, with nearly all neurons (including decision-making Kenyon cells) becoming active by the second step (Fig.~\ref{fig:figure_2}g,i). By contrast, in the biological connectome and in the more biologically plausible randomizations, activity propagates through the network more gradually. As shown in Fig.~\ref{fig:figure_2}f, these connectomes contain far fewer long-range synapses, which limits how quickly signals can spread across the brain. As we discuss in what follows, these differences in the propagation speed and spatial reach of activity likely contribute to the visual task performance differences we observe across ensembles.


\subsection{Connectome models can perform different visual tasks}

We train the biological and randomized connectomes as neural networks in which the anatomical wiring is fixed and learning occurs by adjusting the strength of existing synapses. In practice, each synaptic connection is associated with a trainable parameter that rescales the anatomical synapse count between two neurons, maintaining the constraint that the effective weight cannot exceed the observed number of synapses. Neural activity then propagates through the connectome via the message-passing dynamics described in Methods, using $N=3$ propagation steps. After propagation, the activity of the Kenyon-cell population is averaged and passed to a single linear unit that produces the final decision. All connectome variants—biological and randomized—are trained using the same protocol, ensuring that performance differences arise from network structure rather than from differences in optimization. We evaluate the previously described models on three visual tasks: color discrimination, shape discrimination, and numerosity tasks.

\begin{figure*}[tp]
\centering
\includegraphics[width=0.75\textwidth]{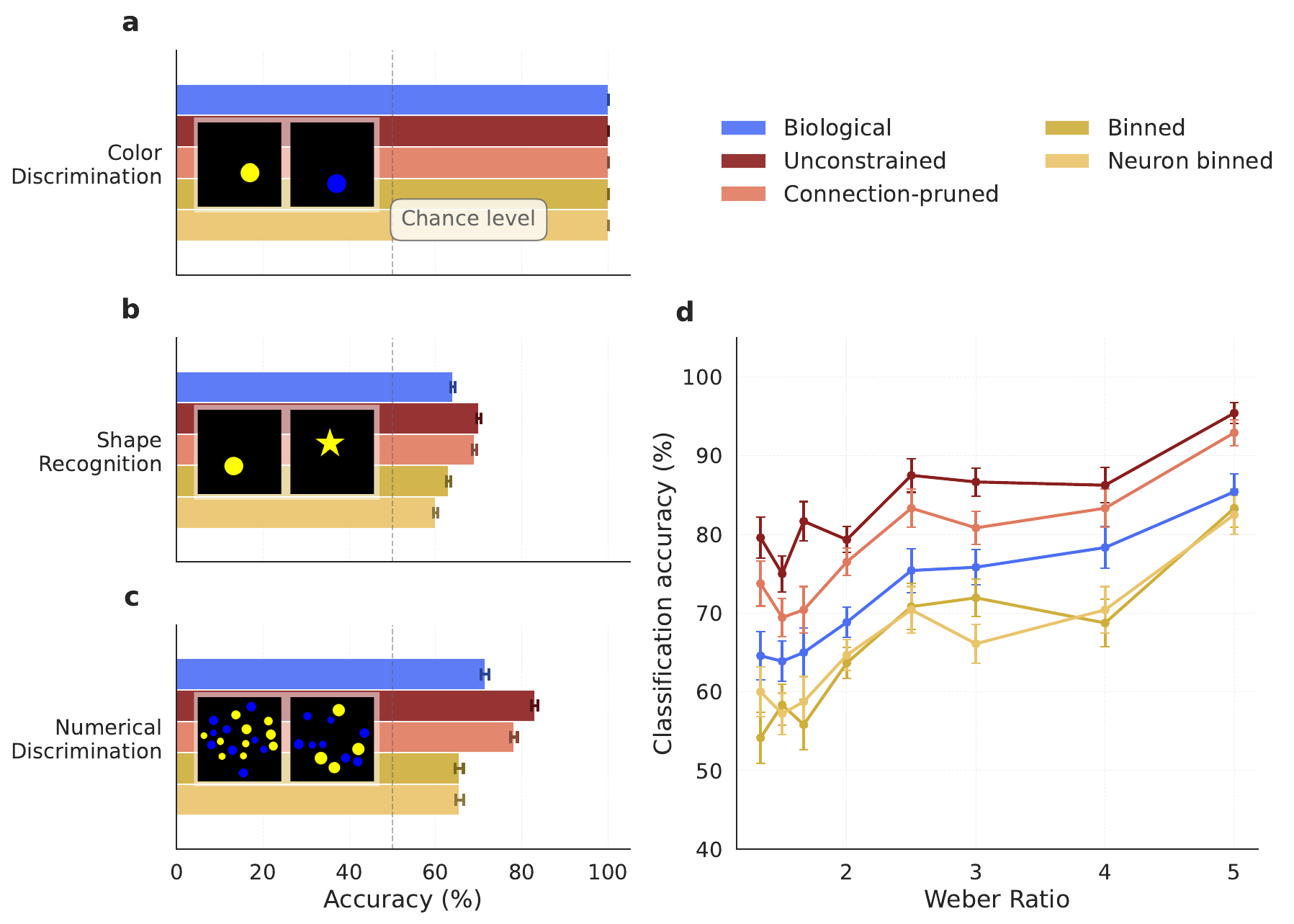}
\caption{\small\textbf{Connectome performance on various visual tasks.}
\textbf{a,} Example stimuli and classification accuracy for color discrimination across the biological connectome and four randomized ensembles; all five models reach 100\% accuracy.
\textbf{b,} Shape-recognition accuracy. The unconstrained and connection-pruned ensembles perform best, followed by the biological, synapse-bin, and neuron-bin models.
\textbf{c,} Numerical-discrimination accuracy, aggregated across the area-controlled held-out images used in \textbf{d}. The unconstrained and connection-pruned ensembles perform best, while the biological connectome outperforms the two distance-binned ensembles.
\textbf{d,} Numerical-discrimination accuracy as a function of the Weber ratio $r=N_{\mathrm{more}}/N_{\mathrm{less}}$. Accuracy generally increases with ratio, showing the ratio-dependent pattern characteristic of approximate number perception. Only the area-controlled condition is shown, in which the total surface area of each color is equalized. 
\textbf{a,b,e} are shown as mean $\pm$ 95\% CI (chance level $=50\%$ for binary tasks). All tasks use  $5\times1{,}000$ test images per class (see Methods).
}
\label{fig:figure_3}
\end{figure*}

\paragraph{Color discrimination.}
Models have to classify images by the color of the circle (blue or yellow) (Fig.~\ref{fig:figure_3}a). Circles can be of different sizes and are placed randomly anywhere in the image. The accuracy of all models, biological or otherwise, is nearly 1 for all the setups that we tried. Indeed, even without training, i.e., from measured synapses alone, the accuracy of the biological connectome was $92\%$.

\paragraph{Shape discrimination.} 
In the shape recognition task, models must classify images containing either circles or stars, with shapes appearing at random positions and with varying sizes (Fig.~\ref{fig:figure_3}b). This task is substantially more difficult than color discrimination; the biological model has $64\%$ accuracy, which increases to $70\%$ for models in the unconstrained ensembles. 

\paragraph{Numerosity.}
Finally, we asked models to perform a numerosity task inspired by Halberda et al.~\cite{halberda_individual_2008} which probes the approximate number system~\cite{dehaene_number_2011} (Fig.~\ref{fig:figure_3}c). This task is relevant because numerical discrimination is a non-trivial visual computation that has been behaviorally demonstrated in \textit{Drosophila melanogaster}~\cite{bengochea_numerical_2023} and provides a canonical assay of approximate number processing. The task consists of choosing the color with the larger number of dots. We consider two stimulus conditions: dots with random sizes without controlling for total color area, and area-controlled arrays in which the total area of the two colors is the same and models cannot use total surface as a cue. Figure~\ref{fig:figure_3}c,d reports the area-controlled condition.

Overall, we find that the average performance across models follows the same pattern as in shape recognition, but with higher accuracy overall. As is standard in psychophysical studies of approximate number discrimination~\cite{feigenson_core_2004, piazza_tuning_2004, halberda_individual_2008}, we also examine performance as a function of the numerical ratio between the two sets, the so-called Weber ratio, $r = N_{\mathrm{more}}/N_{\mathrm{less}}$ where $N_{\mathrm{more}}$ and $N_{\mathrm{less}}$ are the numbers of dots in the more and less numerous colors, respectively (Fig.~\ref{fig:figure_3}d). This allows us to test for the expected Weber-like signature of approximate number perception: discrimination should improve as the ratio between the two numerosities increases.
We find that, across models, accuracy is always above chance and increases with ratio  (for example, for the biological model accuracy increases from $64\%$ at $r=1.5$ to $85\%$ at $r=5.0$), consistent with Weber-like behavior and in line with experiments on real \emph{Drosophila}~\cite{bengochea_numerical_2023}.

\subsection{Wiring constraints limit model accuracy and the biological connectome performs best among biologically-plausible models}

Our results show that task accuracy depends strongly on the wiring constraints imposed on each ensemble (Fig.~\ref{fig:figure_3}). As expected, the unconstrained and connection-pruned connectomes achieve the highest accuracies across tasks. In these ensembles, long-range synapses are more common than in the biological graph, allowing activity to spread farther and reach more neurons in fewer steps. This gives the models access to more neurons for computation, which helps explain their higher accuracy.

By contrast, the synapse-bin and neuron-bin ensembles, which preserve the biological wiring budget and the distribution of synaptic lengths, perform close to but consistently below the biological connectome. Thus, accuracy can be improved beyond biological levels either by relaxing wiring economy (unconstrained) or by changing the distribution of synaptic lengths while keeping total wiring fixed (connection-pruned). However, when models are required to operate under constraints similar to the biological ones and more biologically plausible, the real biological connectome performs best. This suggests that the fly connectome reflects an efficient evolutionary solution---it maximizes accuracy under biological constraints, performing systematically better than nearby random alternatives~\cite{chklovskii_wiring_2002, bullmore_economy_2012}.

\section{Discussion}

A central question in connectomics is to what extent neural computation is already determined by measured synaptic wiring. Here we show that the anatomical connectivity of the \emph{Drosophila} brain, coupled with a biologically faithful model of the compound eye, is sufficient to support several non-trivial visual computations that require a certain level of abstraction. For instance, the connectome-only model we build can identify differences in the number of points of different colors, even though we do not inform the model with the number of points during training.
%
%
This connectome-only design allows us to isolate the computational role of the measured wiring and eye geometry. In contrast to other approaches that augment measured anatomy with theoretical constructs~\cite{lappalainen_connectome_constrained_2023, chen_data_based_2022, shiu_drosophila_2024}, our results indicate that several visual capabilities can already emerge from biologically faithful structure alone. 

Comparing the biological connectome with randomized variants reveals how visual accuracy depends on wiring constraints. When we allow connections to be freely rewired across the brain, ignoring their physical distance, the resulting models achieve the highest accuracies. However, they do so by introducing many long-range connections, which substantially increase both the average connection length and the total wiring cost. 
By contrast, randomizations that are more biologically plausible and preserve both the total wiring budget and the global length distribution yield intermediate performance that is consistently close to, but always below, the biological connectome model. Across tasks, the biological connectome model therefore operates near the optimal accuracy achievable under biologically realistic constraints (Fig.~\ref{fig:figure_3}). Together, these comparisons indicate that although accuracy can be increased by violating biological constraints, the specific pattern of fly connectivity is comparatively efficient when evaluated at its native wiring cost, consistent with principles of wiring economy observed across nervous systems~\cite{chklovskii_wiring_2002, bullmore_economy_2012}.  

Indeed, the existence of ubiquitous long-range connections would have several biological implications. First, they would substantially elevate the energetic cost of maintaining neural activity, owing to increased axonal length, membrane area, and signal propagation demands~\cite{laughlin_energy_2001, niven_energy_2008}, which makes long connection-based architectures metabolically expensive. Second, long-range rewiring also would incur sizable developmental costs because of the need for more complex sets of rules to generate circuit architecture ~\cite{zador_critique_2019, barabasi_genetic_2020} and to provide guidance for axon navigation ~\cite{stoeckli_understanding_2018, sakai_guidance_2012}.  Together, these considerations indicate that although unconstrained wiring can improve task accuracy, it does so by violating energetic, developmental, and informational constraints that shape biological connectomes. Therefore, the biological connectome appears to represent a parsimonious trade-off between computational performance and wiring cost~\cite{chklovskii_wiring_2002, bullmore_economy_2012}.

The randomized connectomes we use provide a principled means of interrogating which features of biological wiring are functionally consequential. By selectively preserving or violating specific anatomical constraints, such comparisons place the biological connectome within a well-defined space of alternative architectures, clarifying not only which computations are supported but also why particular structural features of the connectome may have been favored by evolution. Together, these position connectome-only models as both a benchmark and an explanatory tool for linking neural structure, computation, and biological constraint.

The deliberately simple family of models that we have presented can serve as a stepping stone toward a broader class of connectome-only models, in which the level of neuronal and synaptic description is systematically increased. Starting from the minimal synaptic-gain formulation used here, this framework naturally extends to more biologically and physically faithful connectome-only models in which neurons or synapses are endowed with explicit dynamics, such as graded responses, firing-rate dynamics, relaxation times, or other biophysical properties, in the spirit of classical theoretical neuroscience~\cite{dayan_theoretical_2001, wilson_excitatory_1972, hopfield_neurons_1984}. Nonetheless, our framework sets the foundations to analyze systematically how computations arise from the circuit itself and to relate task performance to patterns of information flow.

All in all, our results delineate the regime in which measured wiring and eye geometry are already sufficient to support non-trivial visual computation. More broadly, they illustrate how connectome-only models can serve as a powerful framework for linking neural structure to function, providing a principled starting point for progressively richer models of brain computation.

\section{Methods}

\subsection{Connectome Data}

We used the proofread adult \emph{Drosophila melanogaster} whole‑brain connectome (version 783) to define the anatomical graph for all experiments~\cite{schlegel_whole-brain_2024}, comprising 139{,}255 proofread neurons and 54.5 million synaptic connections; neuron 3D positions were obtained from the associated dataset~\cite{dorkenwald_neuronal_2024}. Neurons are graph nodes; directed edges represent connected neuron pairs and carry synapse multiplicity. Unless specified, analyses used the right‑eye projection and downstream circuitry.

\subsection{Biologically faithful eye front end}
The eye model converts RGB images into the activation of the ~8,000 retinal photoreceptors present in the connectome while preserving the spatial organization of ommatidia and the spectral tuning of each receptor type via the following steps:

\paragraph{Ommatidial geometry}

We first reconstructed the spatial layout of ommatidia from the connectome data. The 3D coordinates of photoreceptor terminals (R1–R8) were projected onto a two-dimensional plane using principal component analysis (PCA). We then used the positions of the R7 photoreceptors as seeds for a Voronoi tessellation, which defines putative ommatidial catchment regions across the visual field~\cite{wernet_building_2004, paulk_vision_2013} (Fig.~\ref{fig:figure_1}c). Each Voronoi cell therefore represents the angular region sampled by one ommatidium.

For each input image, pixel intensities were averaged within every Voronoi cell to produce a single visual signal per ommatidium. This signal was then delivered to all photoreceptors belonging to that ommatidium. In other words, all receptors within the same ommatidium receive the same spatial input but transform it differently according to their spectral sensitivities (Fig.~\ref{fig:figure_1}d).

\paragraph{Spectral channels}

Photoreceptors were assigned spectral sensitivities according to experimental measurements~\cite{paulk_vision_2013}. Because the input images are RGB and the \emph{Drosophila} vision range extends deep into the UV channel, we shifted the spectral response of all receptors by 200 nm so that they map the RGB channels we can work with. In this way, the UV band was shifted to the blue and thus activated R7 receptors, whereas R8 receptors were divided into two types: R8p shifted from blue to green, and R8y shifted to red.

\paragraph{Photoreceptor activations}

After spectral weighting, each ommatidium produces activations for its eight photoreceptors (R1–R8). These activations form the retinal activity pattern that drives the neural model. Importantly, each photoreceptor is represented as an individual node in the connectome graph, so the final output of the eye model is a vector containing the activation of every retinal neuron.

\subsection{Neural network architecture and propagation}

The connectome is implemented as a message-passing neural network defined on the anatomical graph of the \emph{Drosophila} brain, where each neuron corresponds to a node and each synaptic connection corresponds to a directed edge. The measured connectivity therefore fixes the computational architecture of the model, while neural activity propagates through this graph according to a simple synaptic update rule.

At the beginning of each simulation, retinal neurons are initialized with the photoreceptor activations produced by the eye model, while all other neurons start with zero activity. Activity then propagates through the connectome for $N$ discrete steps (default $N=3$; robustness analyses used $N=2$–$6$). Each propagation step represents one round of synaptic transmission across the network.

The generic equation for the message-propagation model is

\begin{equation}
x_i^{(k)} = \gamma^{(k)}\left(x_i^{(k-1)}, \bigoplus_{j\in\mathcal{N}(i)} \phi^{(k)}\left(x_i^{(k-1)}, x_j^{(k-1)}, e_{ji}\right)\right),
\label{eq:methods_message_passing1}
\end{equation}

where $\bigoplus$ denotes a permutation-invariant aggregation operator, $x_i^{(k)}$ is the activity of neuron $i$ after propagation step $k$, $\mathcal{N}(i)$ denotes the set of presynaptic neurons projecting to $i$, and $e_{ji}$ is the anatomical synapse count from neuron $j$ to neuron $i$. The functions $\phi$ and $\gamma$ respectively represent the synaptic message function and the neuronal activation function.

In our implementation, the aggregation operator is a sum and neurons do not retain their previous state between propagation steps. Instead, each step computes a new activation solely from incoming synaptic inputs, corresponding to neurons firing in response to presynaptic activity and then resetting before the next propagation step. Under these assumptions, Eq.~\ref{eq:methods_message_passing1} reduces to

\begin{equation}
x_i^{(k)} = \gamma^{(k)}\left(\sum_{j\in\mathcal{N}(i)} \phi^{(k)}\left(x_j^{(k-1)}, e_{ji}\right)\right).
\label{eq:methods_message_passing2}
\end{equation}

We further assume a linear synaptic message function in which the contribution of neuron $j$ is proportional to both its activity and the number of anatomical synapses between the two neurons. This yields the update rule

\begin{equation}
x_i^{(k)} =
\gamma^{(k)}\!\left(
\sum_{j\in\mathcal{N}(i)} x_j^{(k-1)}\, e_{ji}\,\omega_{ji} - \xi_i
\right),
\label{eq:methods_message_passing3}
\end{equation}

where $\omega_{ji}$ is a learnable synaptic gain and $\xi_i$ represents a neuronal activation threshold. The effective synaptic weight is therefore proportional to the anatomical synapse count. In the main experiments, the activation function $\gamma$ is taken to be $\tanh$.

Synaptic gains are parameterized as

\begin{equation}
\omega_{ji} = \tanh(\theta_{ji}),
\end{equation}

so that the learned parameters $\theta_{ji}$ produce effective synaptic weights bounded in the interval $[-1,1]$. This allows synapses to become either excitatory or inhibitory during learning while maintaining numerical stability.

\paragraph{Synaptic sign constraints}

In an alternative configuration, we incorporate neurotransmitter annotations from the refined FlyWire dataset~\cite{dorkenwald_neuronal_2024}, which indicate whether each synapse is excitatory or inhibitory. In this case synapse counts $e_{ji}$ are initialized with their biological sign, such that $e_{ji}\in\{-1,1\}\times\text{synapse count}$. To preserve this sign during learning, we restrict synaptic gains to the interval $[0,1]$ using a sigmoid transformation. These variants are explored in supplementary experiments but are not used in the main analyses.

Additional architectural variants, including alternative nonlinearities, normalization schemes, and threshold learning, are described in the Supplemental Material~\cite{SupplementalMaterial}.

\subsection{Synaptic and neuronal plasticity}

The formulation in Eq.~\ref{eq:methods_message_passing3} separates two types of parameters that can in principle adapt during learning: synaptic gains $\omega_{ji}$, which modulate the strength of anatomical connections, and neuronal activation thresholds $\xi_i$, which control the excitability of individual neurons. This allows us to explore different plasticity regimes while keeping the underlying connectome architecture fixed.

We therefore considered four training configurations:

(i) \textbf{Classifier-only}: the connectome dynamics remain fixed, and only the parameters of the final linear decision layer are trained.

(ii) \textbf{Edges-only}: one synaptic gain $\theta_{ji}$ is learned for each nonzero directed neuron-pair connection while neuronal thresholds $\xi_i$ remain fixed; the final linear decision layer is also trained.

(iii) \textbf{Thresholds-only}: synaptic gains remain fixed, but neurons learn activation thresholds $\{\xi_i\}$ that modulate their excitability, with one threshold per retained neuron; the final linear decision layer is also trained.

(iv) \textbf{Edges + thresholds}: both synaptic gains and neuronal thresholds are learned simultaneously, together with the final linear decision layer.

Unless otherwise stated, all experiments reported in the main text use the \emph{edges-only} configuration, in which the connectome topology and neuronal properties remain fixed while one gain per nonzero directed connection is optimized during training.

Exploratory variants incorporating neuronal thresholds or transmitter-specific synaptic signs are described in the Supplemental Material~\cite{SupplementalMaterial} but are not used in the principal comparisons between biological and randomized connectomes.

\subsection{Readout locus}

Decisions are read out from the population of Kenyon cells in the mushroom body, which are widely considered the principal decision-making neurons in the \emph{Drosophila} brain~\cite{aso_mushroom_2014}. After $N$ propagation steps, we read out the population activity of Kenyon cells in the mushroom body and use it as the model's final representation of the stimulus. This choice provides a simple and standardized readout layer for the model, while remaining agnostic about the full circuitry underlying action selection in \emph{Drosophila}.

To obtain a behavioral prediction, this activity is passed to a simple linear classifier, so that the model is as biologically constrained as possible. An additional variant for multi-class tasks that introduces $N_{\mathrm{Kenyon}}\!\times\!C$ auxiliary class units is described in the Supplemental Material~\cite{SupplementalMaterial} but is not used in the principal comparisons.

\subsection{Randomised-connectome ensembles}

To assess which aspects of the biological wiring are important for computation, we compared the measured connectome to ensembles of randomized networks with different degrees of structural constraint. Each ensemble preserves some properties of the biological network while randomizing others, allowing us to isolate the role of wiring economy and spatial organization. The ensembles span a spectrum of biological plausibility: from unconstrained rewiring that ignores spatial embedding to highly constrained variants that preserve the distribution of connection lengths globally or even at the level of individual neurons.

All randomizations preserve the neuron set and are evaluated using exactly the same stimuli and train/test splits as the biological connectome. For each randomization strategy, we used one generated graph instance; the exact four graph files used in the analyses are included in the accompanying data record. We therefore consider four randomized graph configurations: "Unconstrained", "Connection-pruned", "Synapse-bin", and "Neuron-bin":

\paragraph{Unconstrained.}
In the "Unconstrained" randomization, we preserve the number of outgoing connections of each neuron but ignore spatial constraints. For every neuron $j$, we randomly reassign the targets of its outgoing connections while keeping its out-degree fixed. The synapse counts $\{s_{ji}\}$ associated with those connections are then redistributed across the newly assigned targets. Because targets are chosen independently of their physical location, this procedure frequently creates long-distance connections that are rare in the biological connectome. As a result, both the total wiring length $L_{\mathrm{tot}}$ and the mean synapse length $\bar d$ become substantially larger than in the biological network.

\paragraph{Connection-pruned.}
The "Connection-pruned" ensemble starts from an instance of the "Unconstrained" randomization and then reduces its wiring cost to match that of the biological connectome. To do so, we iteratively remove entire connections (setting $s_{ji}\leftarrow0$), preferentially eliminating those that contribute most to the total wiring cost $d_{ji}s_{ji}$, until the total wiring length $L_{\mathrm{tot}}$ equals the biological value. Although this pruning restores the total wiring budget, the remaining connections are still, on average, longer than in the biological network, so the mean connection length $\bar d$ remains higher than biological values.

\paragraph{Synapse-bin.}
In the "Synapse-bin" randomization, we preserve the global distribution of connection lengths while randomizing the wiring pattern. We first group all possible neuron pairs according to the Euclidean distance between their somata ($d_{ji}$) into $B$ distance bins (default $B=100$). Synapses are then shuffled only among neuron pairs that fall within the same distance bin while keeping the total number of synapses in each bin fixed. In this way, any pair of neurons within a given distance bin can receive synapses, regardless of whether they were originally connected. This procedure preserves the overall distribution of connection lengths and therefore keeps both the mean synapse length $\bar d$ and the total wiring cost $L_{\mathrm{tot}}$ equal to those of the biological connectome while randomizing which neuron pairs are connected.

\paragraph{Neuron-bin.}
The "Neuron-bin" randomization further constrains the rewiring by preserving the length distribution of outgoing connections for each individual neuron. For every neuron $j$, its outgoing synapses are grouped into bins according to their wiring length, and synapses are shuffled only within each bin of that neuron. This preserves both the in/out-degree of each neuron and the histogram of outgoing synapse lengths. As a result, the large-scale wiring structure of the network is maintained, and only the distribution of synaptic weights between already connected neuron pairs is altered. Among our randomizations, this ensemble therefore represents the most constrained null model short of the observed biological connectome.

\subsection{Wiring-cost metrics}

To quantify the physical cost of both the biological and the random network architectures, we compute wiring-length metrics based on the spatial positions of neurons in the connectome. Following previous work on wiring economy~\cite{chklovskii_wiring_2002, bullmore_economy_2012}, we define the total wiring length as

\begin{equation}
L_{\mathrm{tot}} = \sum_{(j,i)} d_{ji}\,s_{ji},
\end{equation}

where $d_{ji}$ is the Euclidean distance between the somata of neurons $j$ and $i$, and $s_{ji}$ is the number of synapses connecting them ($s_{ji}=e_{ji}$).

We also compute the synapse-weighted mean connection length,

\begin{equation}
\bar d =
\frac{\sum_{(j,i)} d_{ji}\,s_{ji}}
{\sum_{(j,i)} s_{ji}},
\end{equation}

which measures the average spatial extent of synaptic connections in the network. Synapse-length histograms are annotated with both $L_{\mathrm{tot}}$ and $\bar d$ (Fig.~\ref{fig:figure_2}\textbf{f}). Together, these metrics quantify the wiring economy of each connectome configuration.

\subsection{Computational tasks and stimuli}

We evaluated the models on three visual tasks using synthetic $512\times512$ images rendered according to the eye’s retinotopic sampling (Fig.~\ref{fig:figure_3}a--c). For each task we generated 5{,}000 training, 5{,}000 validation, and 5{,}000 test images per class. Stimuli were designed so that task-relevant information could not be solved through trivial visual cues such as position or total surface area.

\paragraph{Color discrimination.}

In this task the model performs a binary classification between yellow and blue stimuli. Shapes were matched in total area and center position to prevent the model from using simple geometric cues. To test position robustness, we also applied the same retinotopic-sector generalization protocol used in the shape task, in which training and test stimuli are presented in different regions of the visual field.

\paragraph{Shape recognition.}
In this task the model classifies shapes as either circles or stars. Shapes were matched in total area and centre position so that classification could not rely on simple size or location cues. To test whether the models learned abstract shape representations rather than retinotopic matching, we trained them using stimuli confined to the left half of the visual field and evaluated them on stimuli appearing only on the right half. This “retinotopic-sector generalisation” protocol tests whether representations are position-invariant. Behavioural experiments have shown that flies display such position-invariant visual recognition~\cite{tang_visual_2004}.

\paragraph{Approximate Number System (ANS).}
This task evaluates numerical discrimination through binary “more-versus-less” judgments between blue and yellow dot arrays. We evaluated Weber ratios $r\in\{4/3,3/2,5/3,2,5/2,3,4,5\}$. The area-controlled held-out set contained, respectively, 240, 360, 240, 600, 240, 360, 240, and 240 images per ensemble at these ratios, for a total of 2,520 images per ensemble. To control for continuous visual cues, half of the stimuli were area-controlled, meaning that the total surface area of each color was equalized so that the models could not rely on total area as a cue. Such controls are commonly used to isolate numerical processing from continuous quantities~\cite{dehaene_number_2011, nieder_adaptive_2020, halberda_individual_2008}. The remaining stimuli had random dot sizes, preventing the model from learning trivial correlations such as “more dots correspond to smaller dots.” In the main text we report results only for the area-controlled condition, which provides the most direct measure of numerical discrimination.

\subsection{Training protocol and regularisation}

All models were trained using cross-entropy loss and AdamW optimization, with an initial learning rate of $3\times10^{-4}$ and a one-cycle learning-rate schedule. Training proceeded for up to 100 epochs. As implemented in the archived training code, early stopping monitored the epoch-level training loss and training accuracy, with a patience of two epochs and a target training accuracy of 0.99. The same training, validation, and test splits were used for the biological connectome and for all randomized graph ensembles.

To mitigate overfitting, we employed several regularization strategies. These included neuron dropout with probability $p_n$, Kenyon-cell dropout with probability $p_k$, and cell-type dropout applied to selected neuron populations. Exact values for these parameters are reported in the corresponding figure captions and configuration files.

\subsection{Propagation and spatial-reach metrics}

To characterize how activity spreads through the connectome (Fig.~\ref{fig:figure_2}), we computed several propagation metrics during message passing. The fraction of active neurons at step $k$ is defined as the proportion of nodes whose updated activation exceeds a small tolerance $\varepsilon$ (repository default). The fraction of active Kenyon cells is defined analogously, restricting the calculation to neurons in the Kenyon-cell population.

To quantify the spatial extent of signal propagation, we also compute the \emph{mean distance from input} at step $k$, defined as the synapse-weighted mean Euclidean distance between active neurons and the set of retinal input neurons. All metrics are computed under the default training configuration and averaged across connectome ensembles.

\subsection{Software and reproducibility}

The connectome model was implemented as a graph neural network using PyTorch Geometric~\cite{fey_fast_2019} and PyTorch~\cite{paszke_pytorch_2019}. To facilitate reuse and reproducibility, the codebase is organized into three repositories. The core machine-learning model of the fruit fly connectome is provided at~\url{https://github.com/eudald-seeslab/train-your-fly/tree/733a8bdb80cb68089d63f8af803a6180c0c67405} and can be used as a standalone package. The study-specific training, analysis, and figure-generation pipeline is available at~\url{https://github.com/eudald-seeslab/connectome/tree/93476a27692cc5a2e8e4c0ea0f1ec398ab5ae50d}. The stimulus-generation and training-data tools are available at~\url{https://github.com/eudald-seeslab/cogstim/tree/3979e2b0377a17df329fbdb369f0012c23d7c222}.

\begin{acknowledgments}
Funded by the European Union. Views and opinions expressed are, however, those of the authors only and do not necessarily reflect those of the European Union or the European Research Council Executive Agency. Neither the European Union nor the granting authority can be held responsible for them. This work is supported by ERC grant CircuitEvolution (ERC-2024-SyG 101167460); projects PID2022-142600NB-I00 and PID2025-176129NB-I00 from MCIN/AEI/10.13039/501100011033; and by the Government of Catalonia (2021SGR-633 and Industrial Doctorate No. DI-128).
\end{acknowledgments}

\section*{Author contributions}

ECF wrote all the code and performed all the experiments. All authors designed the research, analyzed the results, and wrote the paper.

\section*{Competing interests}

The authors declare no conflict of interest.

\section*{Data availability}

The primary connectivity data used in this study are from the FlyWire FAFB v783 whole-brain connectome release~\cite{flywire_connectivity_2024}. The neuron annotations correspond to FlyWire annotations release v2.1.0, available at~\url{https://github.com/flyconnectome/flywire_annotations/releases/tag/v2.1.0}. Supporting processed biological and randomized connectome graphs, the retinal-neuron coordinate mapping, trial-level prediction files, and numerical-discrimination source-data tables are available on Zenodo~\cite{correigfraga_connectome_data_2026} under a CC BY 4.0 license. No human or sensitive data were used.

\section*{Code availability}

The custom code used in this study is publicly available without access restrictions. The exact version of the study-specific training, analysis, and figure-generation pipeline is available at~\url{https://github.com/eudald-seeslab/connectome/tree/93476a27692cc5a2e8e4c0ea0f1ec398ab5ae50d}. The exact version of the core connectome model and training infrastructure is available at~\url{https://github.com/eudald-seeslab/train-your-fly/tree/733a8bdb80cb68089d63f8af803a6180c0c67405}. The stimulus-generation tools are available at~\url{https://github.com/eudald-seeslab/cogstim/tree/3979e2b0377a17df329fbdb369f0012c23d7c222}.

\bibliography{connectome}

\end{document}